\documentclass[a4paper,11pt]{article}
\pdfoutput=1 

\usepackage{jcappub} 

\usepackage{color}
\usepackage{enumitem}

\title{Comparison and contrast of test-particle and numerical-relativity waveform templates}

\author{J.~F.~Rodr\'iguez}
\author{J.~A.~Rueda}
\author{and R.~Ruffini}
\affiliation{%
ICRANet, Piazza della Repubblica 10, I--65122 Pescara, Italy\\
Dipartimento di Fisica and ICRA, Sapienza Universit\`a di Roma, P.le Aldo Moro 5, I--00185 Rome, Italy
}%
\emailAdd{jose.rodriguez@icranet.org}
\emailAdd{jorge.rueda@icra.it}
\emailAdd{ruffini@icra.it}

\date{\today}

\abstract{
We compare and contrast the emission of gravitational waves and waveforms for the recently established ``helicoidal-drifting-sequence'' of a test particle around a Kerr black hole with the publicly available waveform templates of numerical-relativity. The merger of two black holes of comparable mass are considered. We outline a final smooth merging of the test particle into the final Kerr black hole. We find a surprising and unexpected agreement between the two treatments if we adopt, for the mass of the particle and the Kerr black hole a Newtonian-center-of-mass description, and for the Kerr black hole spin an effective value whose nature remains to be clarified.
}

\begin{document}
\maketitle
\flushbottom

\section{Introduction}\label{sec:1}

Following our recent determination of the ``helicoidal-drifting-sequence'' (hereafter HDS) of a test particle around a Kerr black hole, we attempt to understand the methodology followed in the construction of the waveform templates publicly available from numerical-relativity simulations. For this scope we shall make use of the waveforms of the SXS catalog (see \cite{SXS:catalog} for details), obtained from simulations of binary black-hole coalescences with the Spectral Einstein Code (SpEC).

We adopt the following three steps, which we do not consider to be necessarily valid, adopted only for the sake of a working hypothesis:
\begin{enumerate}
\item 
We adopt the HDS for a test particle following our previous work (\cite{2017arXiv170606440R}).
\item
In order to step from a test particle of mass $m$ to a binary black hole of comparable masses $m_1$ and $m_2$, we adopt the Newtonian-center-of-mass description assuming $m = \mu$, where $\mu = m_1 m_2/M$ is the reduced mass of the binary and $M=m_1+m_2$ is the total binary mass.
\item
We compute the waveform of the HDS all the way up to the passage over the last circular orbit, as originally introduced by \cite{1971ESRSP..52...45R}. As shown in \cite{2017arXiv170606440R}, we do not expect any significant gravitational-wave emission during the final smooth merging of the particle into the Kerr black hole.
\end{enumerate}

\section{General considerations of the reduced mass}\label{sec:2}

It is clear already from the Newtonian-center-of-mass approach that the dimensionless spin parameter, $a/M$, where $a = J/M = \mu \sqrt{r/M}$ is the orbital angular momentum per unit mass and $r$ is the objects interdistance, is typically much larger than unity in any self-gravitating binary system of comparable masses. For equal-masses, it converges from above, namely from $a/M >1$, to $a/M = 1$ only when $r = 16 M$. This implies that only massive neutron stars or black holes can reach black hole formation in their final merger process. Most important, the condition of $a/M = 1$ under these conditions can only by reach from above, i.e. from $a/M >1$, and not by accretion. This gives a tangible way to see implemented, by emission of gravitational waves, the no-hair theorem (see, e.g., Fig. 1 in \cite{1971PhT....24a..30R}).

We consider in our previous work (\cite{2017arXiv170606440R}) the gravitational-wave emission of a test particle, initially in circular orbit around a Schwarzschild or a Kerr black hole. We include the gravitational-wave radiation-reaction into the equations of motion to compute the dynamical evolution of the particle, obtaining the HDS of orbits all the way up to the final smooth merger of the particle into the black hole.

Having obtained this result we here proceed, following similar attempts in the literature, to adopt the HDS of the test particle as an effective body to describe the merger of two black holes of comparable mass. We can not refrain from expressing our surprise for the agreement of this treatment with the one of numerical-relativity waveform templates. 

\section{Test-particle waveform}\label{sec:3}

During the initial phases of the HDS of the particle, the motion is quasi-circular, namely the radial velocity is relatively small with respect to the tangential velocity. Thus, the gravitational waveform can be, in first approximation, constructed from circular-orbit waves. The gravitational wave can be obtained from the scalar $\psi_4$ as \cite{1978ApJ...225..687D}:
\begin{equation}
	\frac{1}{2}\bigl( h_+ -i h_{\times}\bigr) =-\frac{1}{R}\sum_{l,m}\frac{Z_{lm}^H}{\omega_m^2} {}_{-2}S_{lm}(\Theta)e^{i m\Phi}e^{-i\omega_m(t - R^*)},\label{eqn:circular-orbit}
\end{equation}
where $R$ is the distance from the Kerr black hole to the observer, $\Theta$ is the angle between the axis of rotation and the observer, $\Phi$ is the azimuthal coordinate of the orbiting body at $t=0$; $R^*$ is the Kerr ``tortoise'' coordinate, and ${}_{-2}S_{lm}$ are the spheroidal harmonics of spin $-2$ \cite{1973ApJ...185..635T}. The complex numbers $Z_{lm\omega}^H$ that depend on $\omega_m = m \omega$, where $\omega$ is the orbital angular velocity, where computed in \cite{2017arXiv170606440R} to estimate the gravitational-wave radiation flux, $dE/dt$, for a particle moving in a circular orbit on the Kerr metric. This radiation has been computed in the Teukolsky's formalism of curvature perturbations \cite{1973ApJ...185..635T,1974ApJ...193..443T} with the aid of the Sasaki-Nakamura radial equation \cite{1982PThPh..67.1788S,1982PhLA...89...68S}.

As the HDS of orbits progresses, the wave frequency changes with time and we evaluate the acquisition of radial momentum. This implies that the complex number $Z_{lm\omega}$ evolves with time, inducing a variable wave amplitude and phase shift. We also replace $\omega_m(t - R^*)$ in the exponential by $m \phi(t - R^*)$ (see, e.g., \cite{2001PhRvD..64f4004H}), where $\phi$ is the azimuthal coordinate of the test particle along the trajectory. We compute the particle's trajectory following our previous work (\cite{2017arXiv170606440R}) from the equations of motion associated with the Hamiltonian (see, e.g., \cite{1992AnPhy.215....1J}, and references therein)
\begin{equation}\label{eq:H}
H = - P_t = - N^i P_i +  N \sqrt{m^2+\gamma^{i j} P_i P_j},
\end{equation}
where $N = 1/\sqrt{-g^{00}}$, $N^i = - g^{t i}/g^{tt}$ and $\gamma^{i j} = g^{ij} + N^i N^j/N^2 = g^{i j}-g^{t i} g^{t j}/g^{tt}$. The Latin index stands for the spatial coordinates $(r,\theta,\phi)$ and $P_i$ are the spatial momenta. The equations of motion on the equatorial plane $\theta=\pi/2$ are (see \cite{2017arXiv170606440R} for details)
\begin{eqnarray}
\frac{dr}{dt} &=&  \frac{\partial H}{\partial P_r},\qquad \frac{d\phi}{dt} = \frac{\partial H}{\partial P_\phi},\label{eq:rdot}\\
%
%
%
\frac{d P_r}{dt} &=& -\frac{\partial H}{\partial r},\qquad \frac{d P_\phi}{dt} = -\frac{1}{\omega}\frac{dE}{dt}.\label{eq:pphidot}
\end{eqnarray}
As usual we decompose the waveform into the spin-weighted spherical harmonics ${}_sY_{lm}(\theta, \phi)$ as follows \cite{1966JMP.....7..863N}:
\begin{equation}
	R(h_+ - i h_{\times}) = \sum_{l,m} h_{lm}(t-R^*){}_{-2}Y_{lm}(\Theta, \Phi),
\end{equation}
where 
\begin{equation}
h_{lm} = -2\frac{Q_{lm}}{\omega_m^2}e^{-i m \phi(t-R^*)},\quad Q_{lm} = \int d(\cos\Theta)\sum_{l',m'} Z_{l'm\omega}^H{}_{-2}S_{l'm'\omega}(\Theta){}_{-2}Y_{lm}(\Theta, 0). \label{eqn:hlmTP}
\end{equation}

Near the last circular orbit the radial momentum significantly grows (see Fig.~2 in \cite{2017arXiv170606440R}). We include the radial motion effects only implicitly through the orbital phase $\phi$ which is obtained from the numerical integration of Eqs.~(\ref{eq:rdot})--(\ref{eq:pphidot}) which include the effects of both radial drift and radiation-reaction.
%

\section{Comparison of the waveforms}\label{sec:4}

In order to do the comparison of the treatments, we start the HDS at some large distance $r_0$ at time $t=0$ and compute the evolution up to the passage over the last circular orbit, at the time $t=t_{\rm plunge}$. Then, we construct the waveform using the method described in the above section. Since the values of the initial time and phase of the two simulations are arbitrary, we perform a constant change of time and phase which minimizes the overall differences between the two waveforms. Since the comparison is in the waveform at infinity, we assume that the two waveforms are expressed as a function of the same time coordinate.

To quantify the difference between two waveforms one can compute the so-called fitting factor
\begin{equation}
F \equiv (h_1|h_2)/\sqrt{(h_1|h_1)(h_2|h_2)},\quad (h_1|h_2) \equiv 4 \mathfrak{Re}\left[\int_{0}^{\infty} h_1(f)\tilde{h}_2(f)/S_n(f)df\right],
\end{equation}
where $f$ is the gravitational-wave frequency in the detector's frame, $\tilde{h}_i(f)$ is the Fourier transform of the waveform $h_i(t)$ and $S_n(f)$ is the power-spectrum density of the detector's noise. For the latter we use the Advanced LIGO noise (see, e.g., \cite{2016LRR....19....1A}). Through the fitting factor one can also define the so-called mismatch, ${\cal M} \equiv 1-F$. Since the function $S_n(f)$ is given in physical units (Hertz) then a value for the total mass of the system has to be specified to calculate the fitting factor. For all the examples shown below we set $M = 70~M_\odot$.

Another way to quantify the difference between two waveforms is by the \emph{intrinsic} time-domain phase evolution $Q_\omega= \omega^2/\dot{\omega}$, where $\omega = d \phi/dt$ and $\phi$ is the gravitational-wave phase. We did not take into account \emph{the small correction in the phase} from the term $Q_{lm}$ in Eq.~\eqref{eqn:hlmTP} to avoid the noise arising from the interpolation of the radiation flux. To calculate $Q_\omega$ we proceed as in Ref.~\cite{2013PhRvD..87h4035D}, although some difficulties were reported there due to the inherent oscillations present in the numerical-relativity data. We prefer here to not perform any fit of the function $Q_\omega$ of the numerical-relativity simulation.

\subsection{Merging black holes of equal-mass and equal, aligned spins}\label{sec:4.1}

We turn now to compare and contrast some waveforms of the SXS catalog with the ones obtained from the HDS treatment of the present work. First, we focus on simulations of binaries of merging black-holes with  equal-mass and equal, aligned spins.

\begin{figure*}
\centering
\includegraphics[width=\hsize,clip]{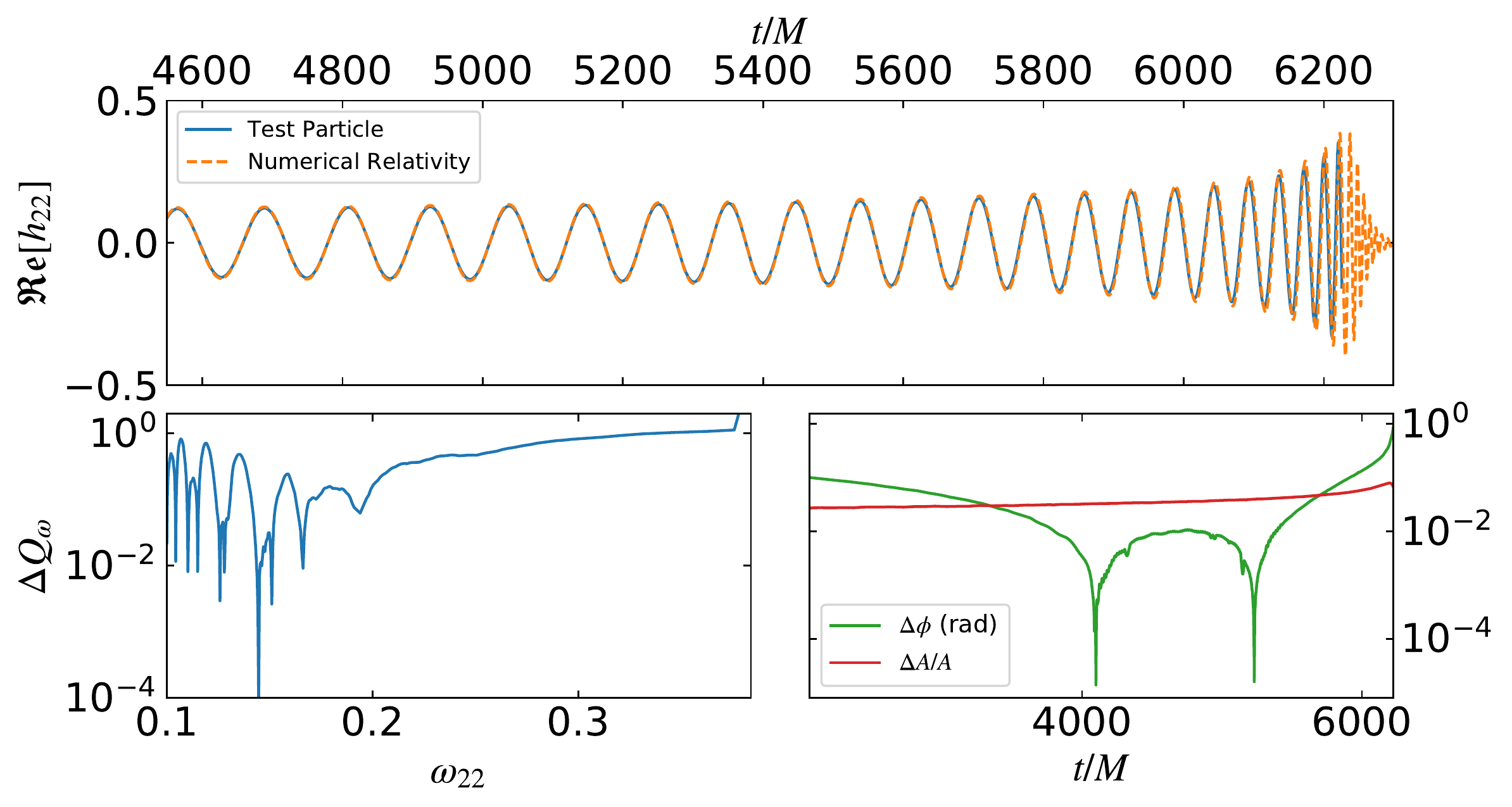}
\caption{Comparison of the HDS and the numerical-relativity waveforms. Top panel: the dashed orange curve is the numerical-relativity waveform \texttt{BBH:0230} \cite{SXS:catalog,Chu:2015kft} of the coalescence of a binary black-hole with $m_1=m_2$ and $a_1/m_1=a_2/m_2=0.8$, forming a Kerr black hole with spin parameter $a_f/M_f=0.907516$. The continuous blue curve is the test-particle waveform during the HDS adopting $m = M/4$ and a dimensionless spin of the Kerr black hole nearly equal to the one of the newly-formed Kerr black hole of the merger simulation, $0.9075$. The time is normalized to the total binary mass, $M$, and the comparison is made up to the instant of passage of the test particle at the location of the last stable circular orbit. For the sake of clarity we show here the waveforms in the last part of the evolution. Left lower panel: intrinsic time-domain phase difference evolution $\Delta Q_{\omega} = |Q_{\omega}^{\textrm{TP}} - Q_{\omega}^{\textrm{NR}}|$ as a function of the frequency of the $l=m=2$ gravitational-wave mode, $\omega_{22}$. We recall that, in this regime of the inspiraling $\omega_{22} \approx 2 \omega$, where $\omega$ is the orbital angular velocity. Right lower panel: phase difference $\Delta \phi$  (in radians; green curve) and relative difference of the amplitudes of the two waveforms shown in the top panel during the entire time of the comparison.}\label{fig:waveform09}
\end{figure*}

We start the comparison with the numerical-relativity simulation \texttt{BBH:0230} \cite{SXS:catalog,Chu:2015kft}: the coalescence of two black holes with $m_1=m_2 = M/2$ and dimensionless spin parameters $a_1/m_1=a_2/m_2=0.8$, forming a Kerr black hole with dimensionless spin parameter $a_f/M_f = 0.907516$. This system is particularly interesting since it is characterized by equal-mass and high-spin components, properties which are in principle different from the non-spinning, test particle domain which we adopt here. No agreement between the two treatments should be a priori expected.

Fig.~\ref{fig:waveform09} shows the comparison of the two waveforms, the one of numerical-relativity simulations with the one of a test particle during the HDS around a Kerr black hole. We adopt a test particle of mass $m = \mu = m_1 m_2/M = M/4$ and a Kerr black hole with spin parameter nearly equal to the one of the newly-formed Kerr black hole in the merger, i.e. $0.9075$. For completeness of the comparison we also show, for this time-interval, the intrinsic time-domain phase evolution $Q_{\omega}$, and the difference between the gravitational-wave phases, $\Delta \phi$ (green curve), and the relative difference between the waveform amplitudes, $\Delta A/A$ (red curve).

We obtain for the waveforms of Fig.~\ref{fig:waveform09} a value $F\approx 0.993$, so a mismatch ${\cal M} = 0.007$, during the entire time-interval of the comparison, i.e.~$t/M \approx 1702.03$--$6182.19$, corresponding to an interval of separation distances $r/M= 14.95$--$2.27$, where the latter is the location of the last circular orbit. It can be seen that, regardless of the $Q_{\omega}$ oscillations for the numerical-relativity data, $\Delta Q_{\omega} = |Q_{\omega}^{\textrm{TP}} - Q_{\omega}^{\textrm{NR}}| \lessapprox 1 $.

\begin{figure}
    \centering
        \includegraphics[width=\hsize,clip]{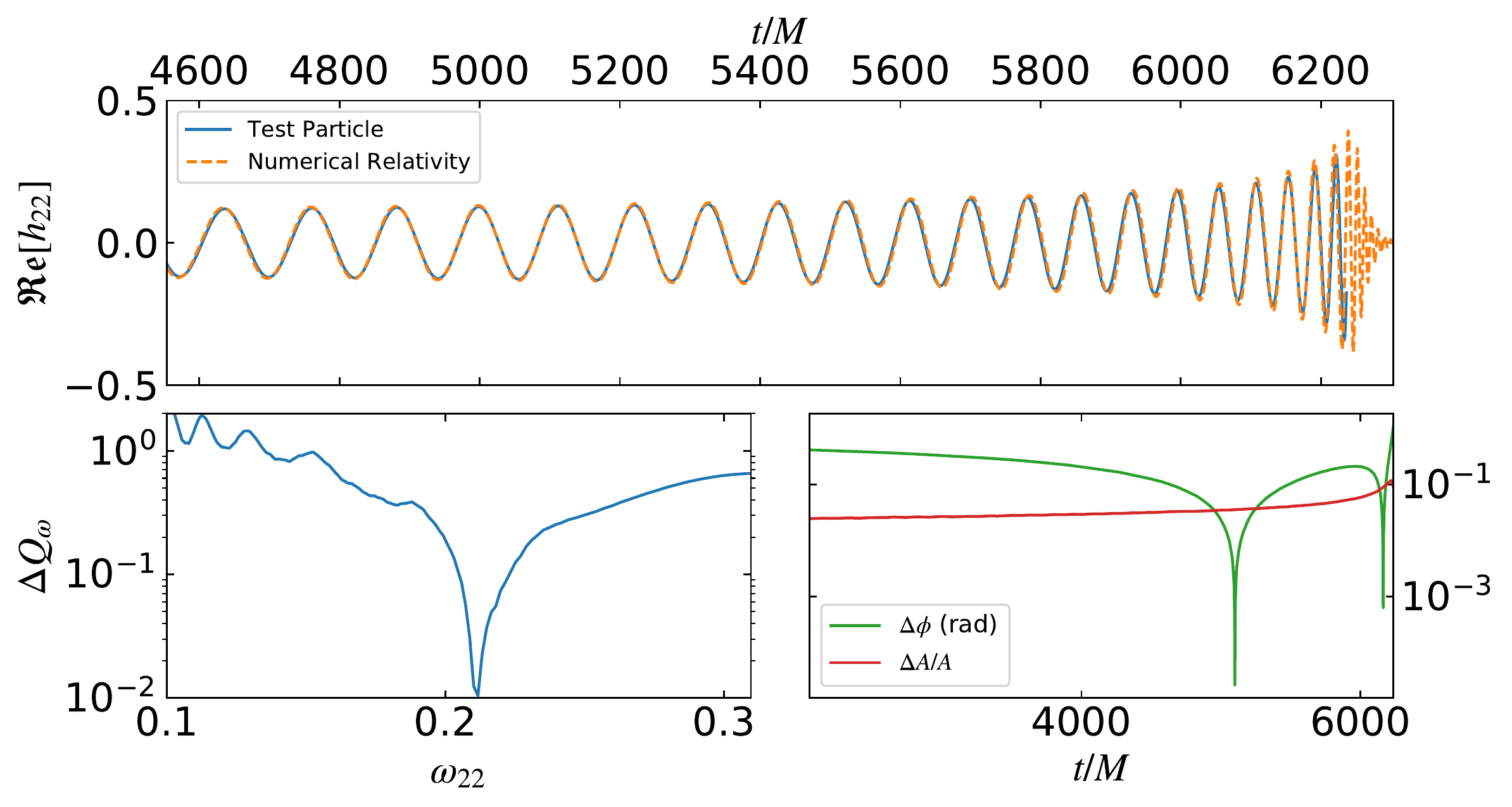}
    \caption{Comparison of the HDS and the numerical-relativity waveforms. Top panel: The dashed orange curve is the numerical-relativity waveform \texttt{BBH:0228} \cite{SXS:catalog,Chu:2015kft} of the coalescence of a binary black-hole with $m_1=m_2$ and $a_1/m_1=a_2/m_2=0.6$, forming a Kerr black hole with spin parameter $a_f/M_f=0.857813$. The continuous blue curve is the test-particle waveform during the HDS adopting $m = M/4$ and a dimensionless spin of the Kerr black hole $a_{\textrm{eff}}/M=0.8$. We use for the HDS the same mass-ratio of the numerical-relativity simulation. Left lower panel: intrinsic time-domain phase difference evolution $\Delta Q_{\omega} = |Q_{\omega}^{\textrm{TP}} - Q_{\omega}^{\textrm{NR}}|$ as a function of the frequency of the $l=m=2$ gravitational-wave mode, $\omega_{22}$.  Right lower panel: phase difference $\Delta \phi$ (in radians; green curve) and relative difference of the amplitudes of the two waveforms.}\label{fig:SXS:228}
\end{figure}

The above agreement between the two waveforms, both in amplitude and in phase, is remarkable and unexpected. We now proceed to the results of additional representative cases. We show in Fig.~\ref{fig:SXS:228} the comparison with the numerical-relativity simulation \texttt{BBH:0228} characterized by two black holes with aligned spins $a_1/m_1=a_2/m_2=0.6$ that merge and form a black hole with spin $a_f/M_f=0.857813$ \cite{SXS:catalog,Chu:2015kft}. We found here a new feature with respect to the previous comparison: the best matching waveform did not correspond to the one generated by the HDS in a Kerr black hole with (nearly) the same spin of the newly-formed black hole. Instead, we found that the agreement is obtained for the HDS in a Kerr black hole with an ``effective'' spin parameter $a_{\textrm{eff}}/M=0.8$. The fitting factor for this case is $F=0.972$.

\begin{figure*}
\centering
        \includegraphics[width=0.49\hsize,clip]{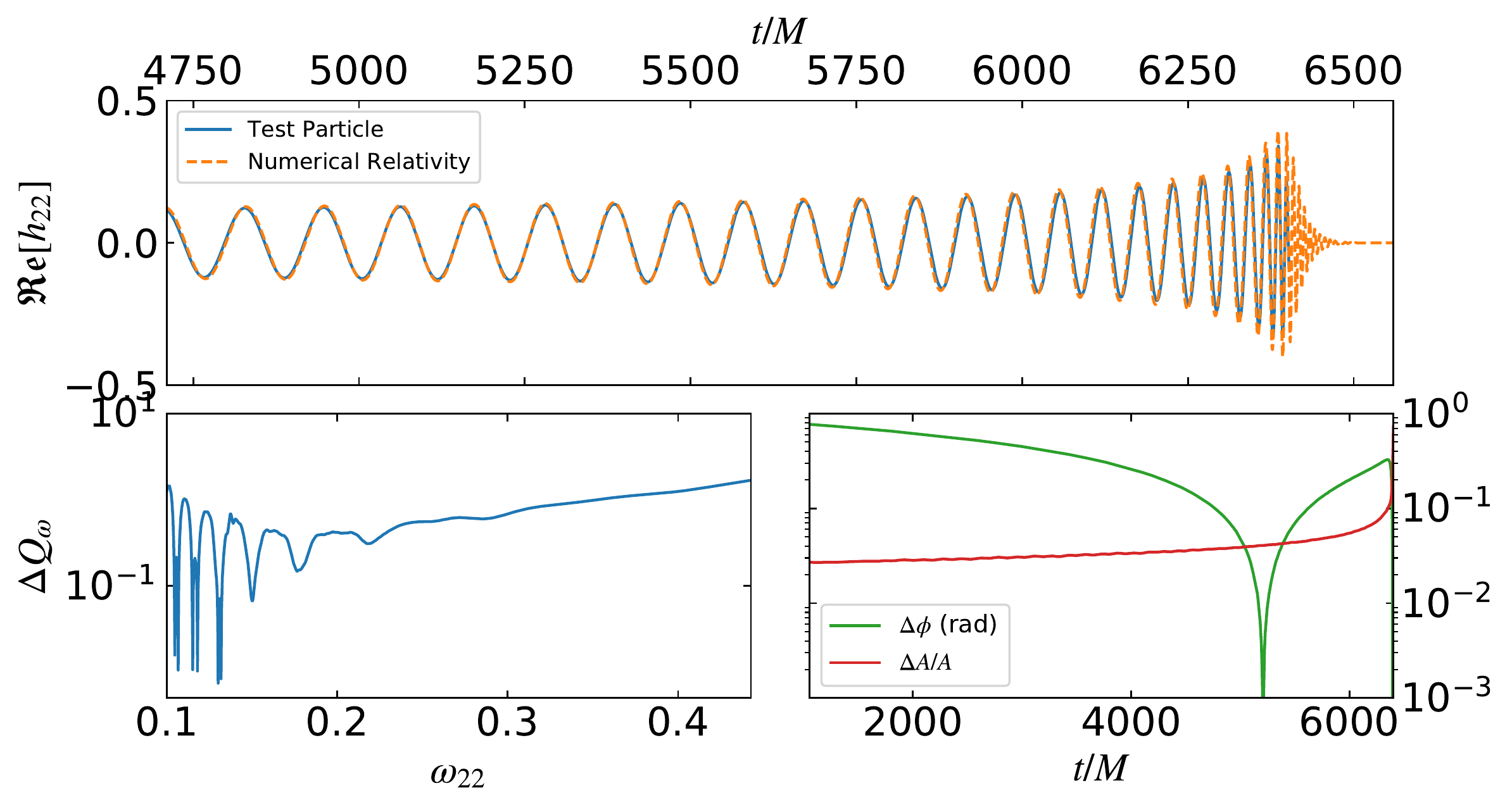}
    	\includegraphics[width=0.49\hsize,clip]{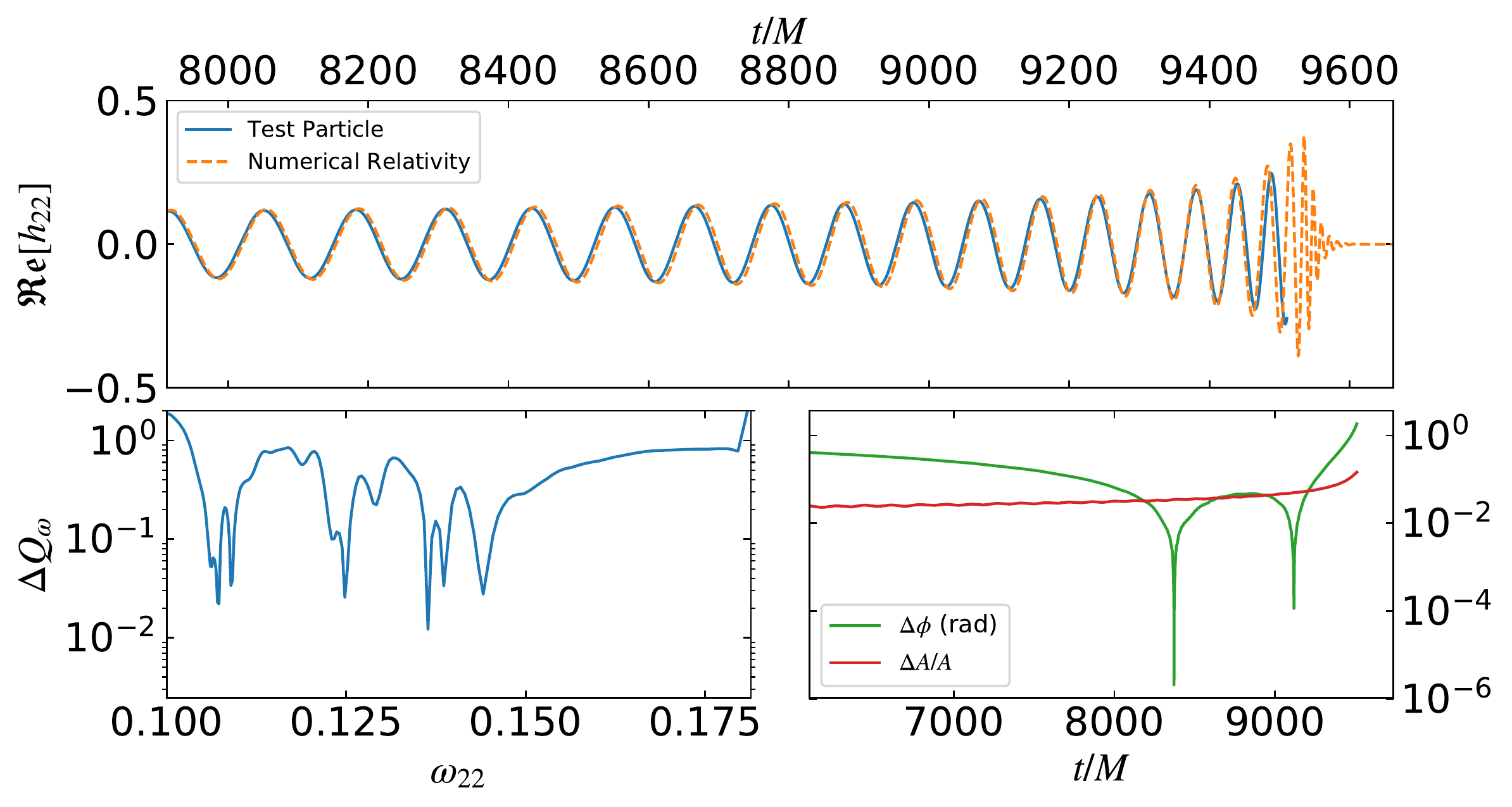}
        \label{fig:SXS:0157-0001}
        \caption{Comparison of numerical-relativity waveform and HDS waveforms. Left panel: \texttt{BBH:0157}, coalescence of a binary black-hole with $m_1=m_2$ and $a_1/m_1=a_2/m_2=0.949586$, forming a Kerr black hole with spin parameter $a_f/M_f=0.940851$ \cite{SXS:catalog,Hemberger:2013hsa}. For the HDS treatment we find an effective spin parameter $a_{\textrm{eff}}/M_f=0.99$. Right panel: \texttt{BBH:0001}, coalescence of a binary black-hole with $m_1=m_2$ and $a_1/m_1=a_2/m_2=1.2\times 10^{-7}$ (i.e.~spinless case), forming a Kerr black hole with spin parameter $a_f/M_f=0.686461$ \cite{SXS:catalog,Mroue:2013xna}. For the HDS treatment we find an effective spin parameter $a_{\textrm{eff}}/M=0.36$. Again, we found that the effective spin parameter of the Kerr black hole in the HDS treatment is neither the one of the newly-formed black hole nor the one of the merging black holes of the numerical-relativity simulation.}
\end{figure*}

The above simulation hinted us the existence of an effective spin parameter for our simulations for which there is a very good matching with the numerical-relativity simulations. Thus, we performed more comparisons with other waveforms of the SXS catalog to confirm it. The results are presented in Fig.~\ref{fig:SXS:0157-0001} and Table~\ref{tab:comparison}. We can conclude from this first part of our analysis that, if we adopt for the HDS treatment the same mass ratio as the one of the numerical-relativity simulation, it can be always found an effective spin of the Kerr black hole of the HDS approach for which the waveforms of the two treatments are in excellent agreement. It can be also seen from $\Delta Q_\omega$ that, regardless of the oscillations inherent in the numerical-relativity simulations, the two phase evolutions agree each other.
Nevertheless, it can be seen that the agreement between the two waveforms decreases in some part of the evolution, suggesting that the effective spin $a_{\textrm{eff}}$ might change with time.

\begin{table}
\centering
        \caption{Column 1: Code of the numerical-relativity simulation of the SXS catalog \cite{SXS:catalog}. Column 2: Spin parameter of the merging black holes, $a_i/m_i$. Column 3: Spin parameter of the newly-formed black hole, $a_f/m_f$. Column 4: Effective spin $a_{\textrm{eff}}/M_f$ of the Kerr black hole in the HDS treatment that gives good agreement with the numerical-relativity simulation. Column 5: Fitting factor between the numerical-relativity and HDS waveforms. All the simulations are for equal-mass binaries.}
    \begin{tabular}{lcccc}
       Simulation & $a_i/m_i$ & $a_f/M_f $ & $a_{\textrm{eff}}/M_f$ & F\\
        \hline
        \texttt{BBH:0001} & $1.209309\times 10^{-7}$ & 0.686461 & 0.36 & 0.96 \\
        \texttt{BBH:0157} & 0.949586 & 0.940851 & 0.99 & 0.93 \\
        \texttt{BBH:0228} & 0.600000 & 0.857813 & 0.80 & 0.972 \\
        \texttt{BBH:0230} & 0.800000 & 0.907516 & 0.9075 & 0.993 \\
        \hline
    \end{tabular}
    \label{tab:comparison}
\end{table}


\subsection{Merging black holes of unequal-mass and spinless}\label{sec:4.2}

\begin{figure*}
\centering
\includegraphics[width=0.49\hsize]{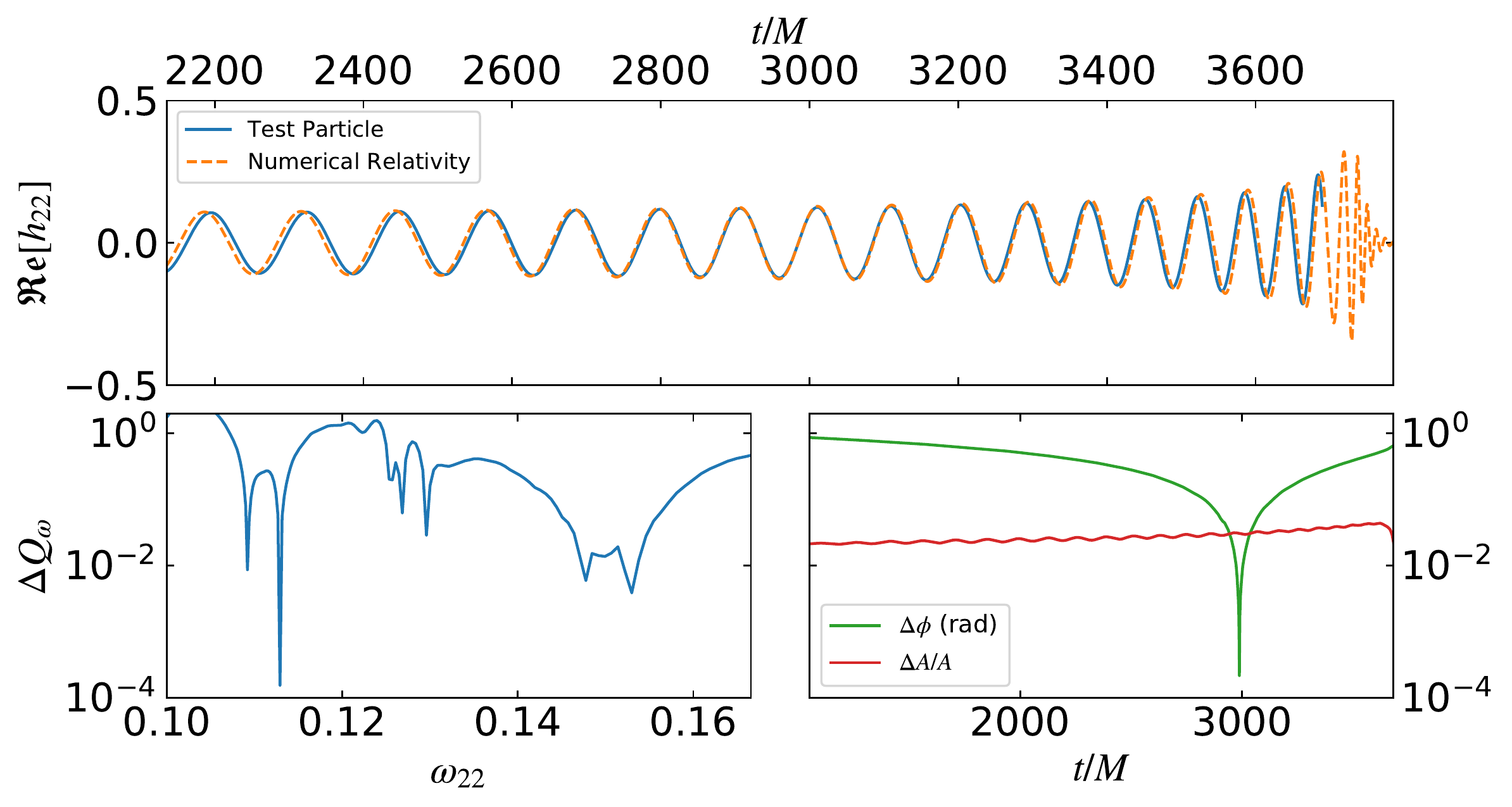}
\includegraphics[width=0.49\hsize]{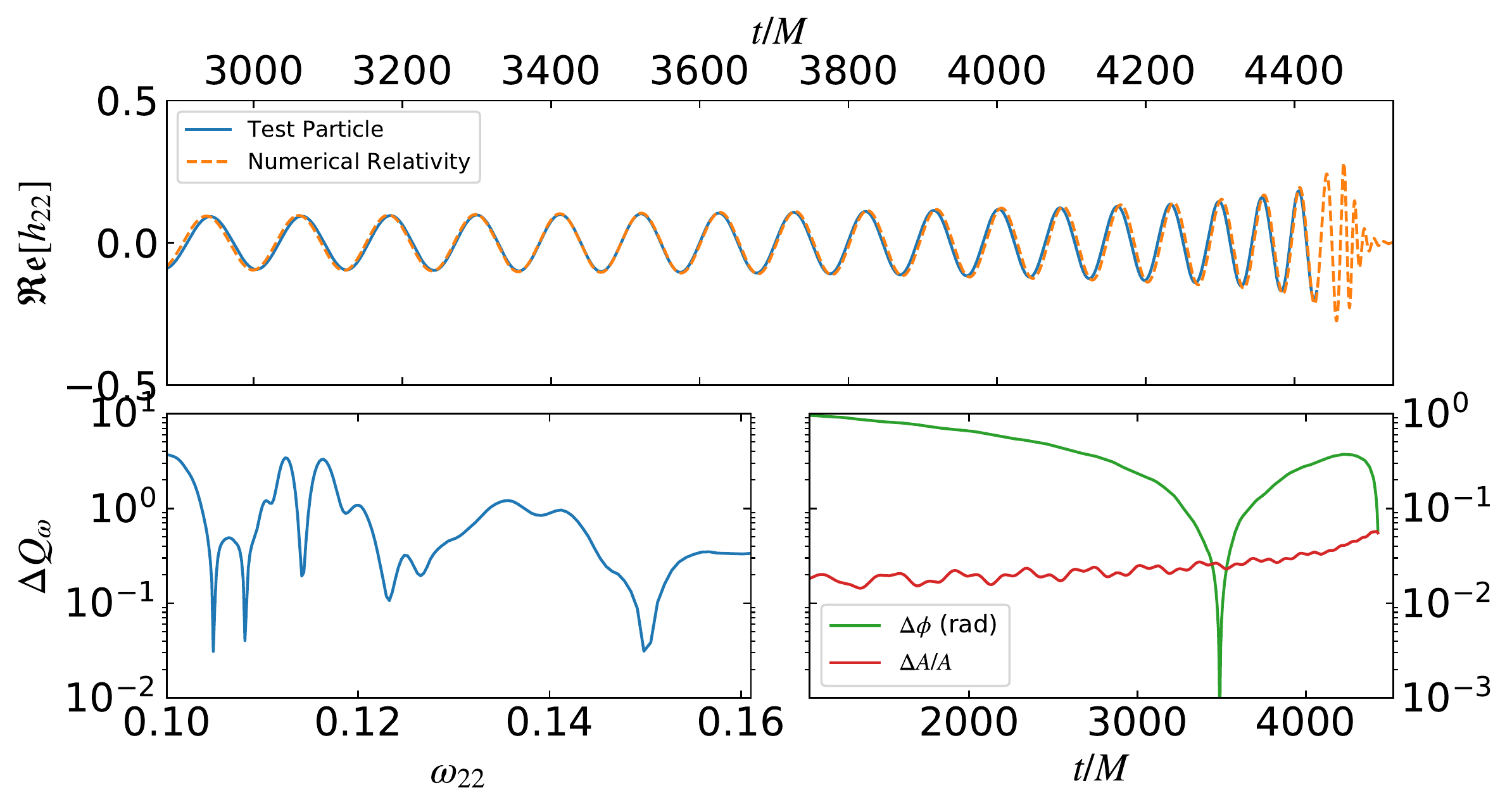}
\includegraphics[width=0.49\hsize]{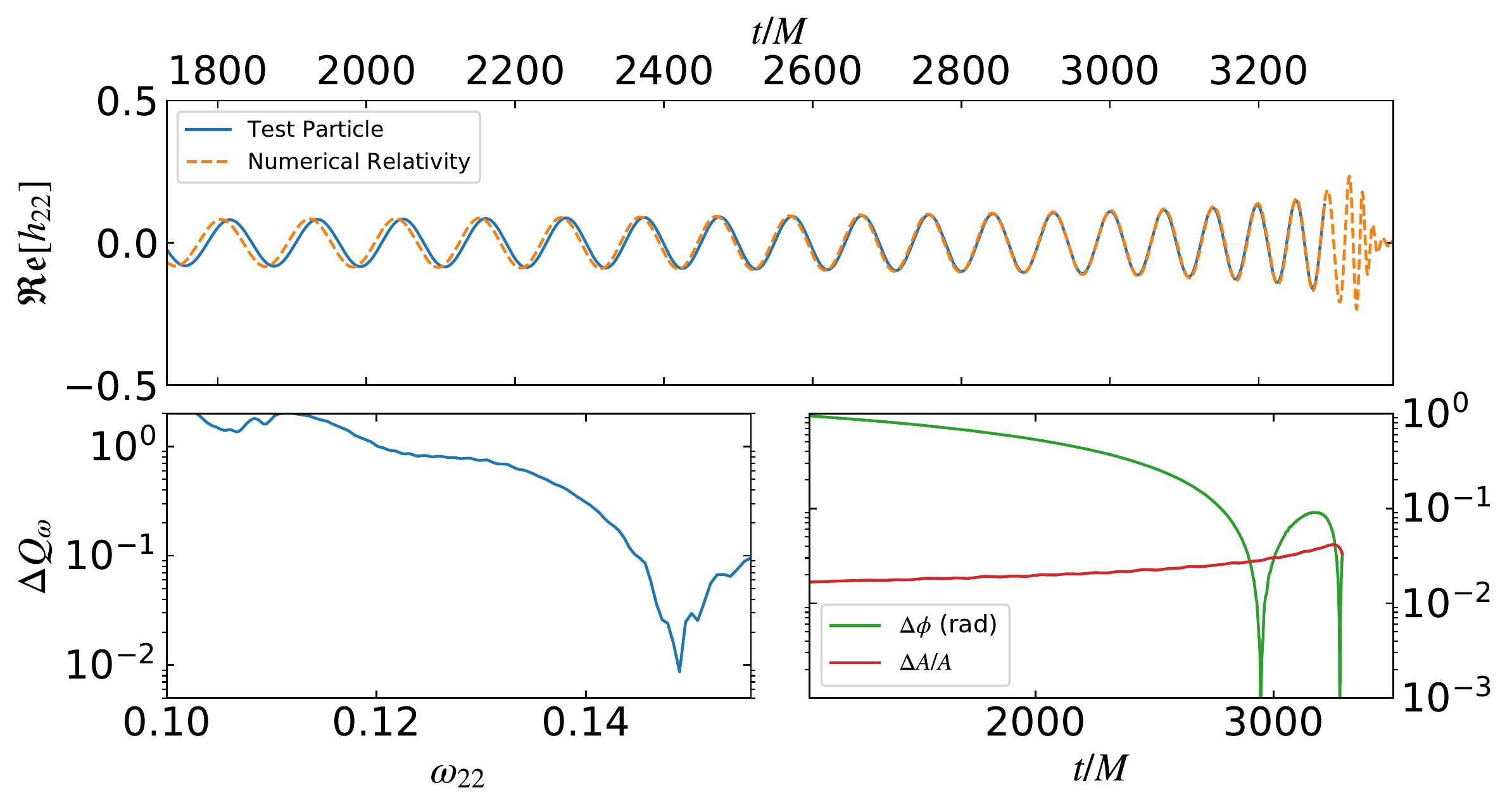}
\caption{Comparison of three numerical-relativity simulations of spinless merging binary black holes with the current HDS approach. Top left panel: \texttt{BBH:0169} numerical-relativity waveform of mass ratio $q=1/2$ \cite{SXS:catalog,Buchman:2012dw}. The effective spin for this case is $a_{\textrm{eff}}/M_f=0.33$ and the fitting factor between the two waveforms is $F=0.95$. Top right panel: \texttt{BBH:0030} numerical-relativity waveform of mass ratio $q=1/3$ \cite{SXS:catalog,Mroue:2013xna}. The effective spin is $a_{\textrm{eff}}/M_f=0.29$ and the fitting factor $F = 0.965$. Bottom panel: \texttt{BBH:0182} numerical-relativity simulation of mass ratio $q=1/4$ \cite{SXS:catalog,Blackman:2015pia}. The effective spin is $a_{\textrm{eff}}/M_f=0.25$ and the fitting factor $F = 0.963$.}
\label{fig:spinless}
\end{figure*}

We also analyzed the case of spinless merging black holes for different mass ratios, $q = m_2/m_1 = 1,1/2,1/3,1/4$, with $m_1 \geq m_2$ (see Fig.~\ref{fig:spinless}). We have found that the effective spin varies proportionally with the binary mass ratio. This also shows that, although the black holes do not spin, there is some spin on the background spacetime even before the formation of the final spinning black hole. The effective spin turned out to be always lower than the spin of the newly-formed black hole.

As we have mentioned, the formation of the Kerr black hole from binaries of black holes of comparable masses occurs from above, namely from $a/M>1$ at larger distances to $a/M \lesssim 1$ as the objects approach each other. Only in the case of binaries with extreme mass-ratios $\mu/M \ll 1$, the merger leads to a slowly-rotating black hole, and, only when $\mu/M\to 0$, the formation of a Schwarzschild black hole can be approached. This is consistent with our results above of an effective spin of the black hole proportional to the mass ratio, so that we expect a vanishing spin, i.e. a Schwarzschild black hole, only in the limiting case of a vanishing mass ratio.

\section{Mass of the newly-formed black hole}\label{sec:5}

We have shown in \cite{2017arXiv170606440R} that no significant gravitational radiation is expected after the passage of the test particle over the last circular orbit. Thus, the mass of the newly-formed black hole is expected to be
\begin{eqnarray}
M_{\rm BH} = M -\Delta E_{\rm rad},\label{eq:Mbh}\\
\Delta E_{\rm rad} = m - H_{\rm plunge}\label{eq:Erad},
\end{eqnarray}
where $H_{\rm plunge} \equiv H (t = t_{\rm plunge})$ is the value of the Hamiltonian (energy) of the particle (\ref{eq:H}) during the final smoothly merging into the black hole (see \cite{2017arXiv170606440R} for details). As we have shown in \cite{2017arXiv170606440R} (see Sec.~V therein), due to the radial drift and the radiation-reaction effects leading to the HDS, when the particle passes over the location of the last circular orbit $H_{\rm plunge}$ is smaller than the energy of a particle in circular orbit in the Kerr geometry:
\begin{equation}
\frac{E}{m} = \frac{r^2 - 2 M r + a M^{1/2} r^{1/2}}{r (r^2 - 3 M r + 2 a M^{1/2} r^{1/2})^{1/2}}, \label{eq:Elco}
\end{equation}
evaluated at the last circular orbit, which facilitates the smooth merging of the particle to the Kerr black hole (see \cite{2017arXiv170606440R} for details).

\begin{figure}
\centering
\includegraphics[width=0.7\hsize,clip]{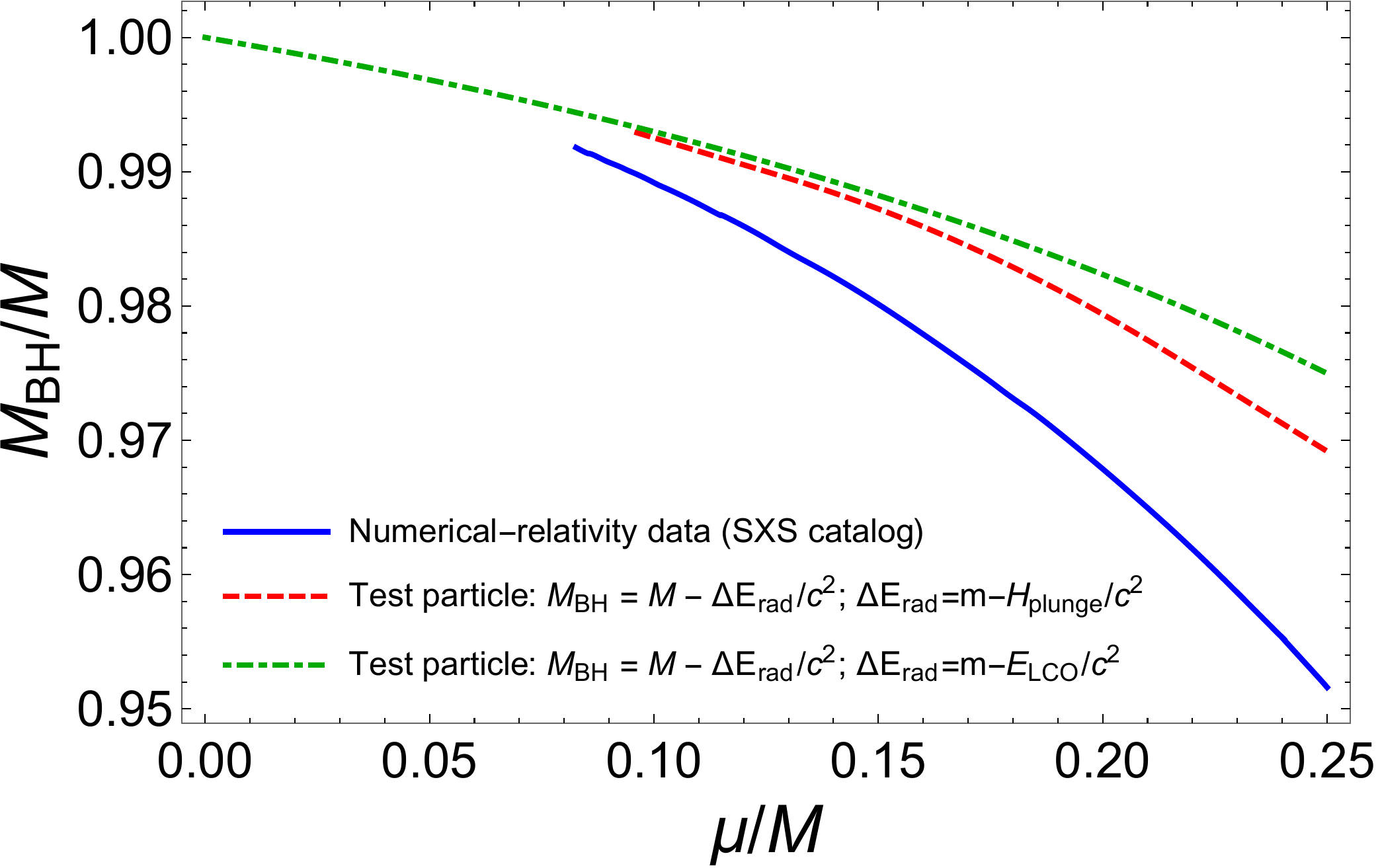}
\caption{Comparison of the mass of the newly-formed black hole predicted by the HDS of the test particle and numerical-relativity simulations of binary black hole mergers. Blue-solid curve: numerical-relativity simulations (\cite{SXS:catalog}) for coalescences of binary black holes with different values of the mass-ratio $q=m_2/m_1$ (we adopt $m_1 \geq m_2$), thus different values of the ratio $\mu=M = q/(1+q)^2$, and non-spinning, i.e. $a_1/m_1=a_2/m_2=0.0$. Red-dashed curve: test particle prediction by Eq.~(\ref{eq:Mbh}) adopting $\Delta E_{\rm rad} = m - H_{\rm plunge}$, Eq.~(\ref{eq:Erad}). Green-dot-dashed curve: test particle prediction by Eq.~(\ref{eq:Mbh}) adopting $\Delta E_{\rm rad} = m - E_{\rm LCO}$ with $E_{\rm LCO}$ given by Eq.~(\ref{eq:Elco}), evaluated at the last circular orbit of the formed Kerr black hole. In our treatment we adopt the mass of the particle, $m$, equal to the binary reduced-mass, $\mu$, and the spin of the Kerr black hole equal to spin of the newly-formed black hole in the merger. These simulations range values of spin parameter from $\approx 0.3$ in for $\mu/M \approx 0.1$ to $\approx 0.7$ for $\mu/M = 0.25$.}\label{fig:Mbh}
\end{figure}

We compare and contrast in Fig.~\ref{fig:Mbh} the mass of the newly-formed black hole predicted by the test particle treatment (\ref{eq:Mbh}) and by the numerical-relativity simulations \cite{SXS:catalog}. The numerical-relativity data in this plot refer to all the available simulations of the waveform catalog of coalescences of binary black holes with non-spinning components. These simulations correspond to different values of the mass-ratio $q=m_2/m_1$ (we adopt $m_1 \geq m_2$), thus different values of the ratio $\mu/M = q/(1+q)^2$. In the case of the above simulations of non-spinning components, the spin parameter of the newly-formed Kerr black hole ranges from $ \approx 0.3$ (for $\mu/M \approx 0.1$) to $\approx 0.7$ (for $\mu/M = 1/4$). We recall that we adopt in the comparison the spin of the Kerr black hole equal to spin of the newly-formed black hole in the numerical-relativity merger simulation.

We find that despite the agreement in the waveforms, the mass of the newly-formed black hole in numerical-relativity simulations is smaller than the one from the HDS of the test particle. It implies the existence in the numerical-relativity simulations of an additional gravitational radiation after the passage of the particle over the last circular orbit, in contrast with the expectations from our results in \cite{2017arXiv170606440R}. At this stage, we do not find any physical reason that explains such an extra loss of gravitational energy at expenses of the black hole mass. We have pointed out in \cite{2017arXiv170606440R} the disagreement between our estimates of the energy radiated with the ones in \cite{2000PhRvD..62l4022O}, which also obtain a much larger radiation from their semi-analytic treatment of the final plunge of the particle into the Kerr black hole. 

We are currently performing numerical calculations of the gravitational radiation from the Teukolsky equation adopting the actual plunge trajectory into the Kerr black hole after the last circular orbit as described in \cite{2017arXiv170606440R}. The results will be presented elsewhere including a waveform analysis.

\section{Conclusions}\label{sec:6}
\begin{enumerate}

\item 
It has been often emphasized in the literature by different groups the necessity of large computing facilities to perform numerical-relativity simulations (see \cite{2015LRR....18....1C} for a recent review, and references therein). However, the results of this article show they can be reproduced with the above theoretical treatment on a single laptop, representing an alternative, more direct, approach.

\item 
Of course this remarkable agreement does not prove the physical veracity of the assumptions we have made but, on the contrary, requires that more input on the assumptions and details of the numerical-relativity simulations be made publicly available, in order to formulate a diagnosis of this unexpected result and a priori unlikely theoretical event.

\item 
The agreement between the two treatments, with and without considering the intrinsic spins of the merging black holes, appears to be due to the dominating value of the binary angular momentum over the one of the individual spins of the merging black holes. What clearly stands from this Letter is a call for attention to the non-applicability of relativistic orbits in the Schwarzschild metric and the neglect of the total binary angular momentum which needs to be taken to the general attention.

\end{enumerate}

\acknowledgments

We had the pleasure during the entire development of this work to discuss in our ICRANet headquarter in Pescara with Prof. Roy Kerr. To him, and to the entire ICRANet Faculty and Staff, goes our gratitude.

\bibliographystyle{JHEP}
\bibliography{references,references_SSX}

\end{document}